# AN 8 GEV LINAC AS THE BOOSTER REPLACEMENT IN THE FERMILAB POWER UPGRADE*

D. Neuffer†, S. Belomestnykh, M. Checchin, D. Johnson, S. Posen,
E. Pozdeyev, V. Pronskikh, A. Saini, N. Solyak, V. Yakovlev
Fermilab, Batavia, IL 60510, USA

*Abstract*

Increasing the Fermilab Main Injector (MI) beam power above ~1.2 MW requires replacement of the 8 GeV Booster by a higher intensity alternative. Earlier, rapid-cycling synchrotron and linac solutions were considered for this purpose. In this paper, we consider the linac version that produces 8 GeV H⁻ beam for injection into the Recycler Ring (RR) or MI The new linac takes ~1 GeV beam from the PIP-II linac and accelerates it to ~ 2 GeV in a 650 MHz SRF linac, and then accelerates to ~8 GeV in an SRF pulsed linac using 1300 MHz cryomodules. The linac components incorporate recent improvements in SRF technology. This Booster Replacement linac (BRL) will increase MI beam power to DUNE to more than 2.5 MW and enable next-generation intensity frontier experiments.

## INTRODUCTION

The PIP-II project will provide an 800 MeV proton beam with CW capability, with beam power up to the MW level available for user experiments [1]. However, the amount of beam that can be transmitted to the Main Injector (MI) is limited by the 0.8—8.0 GeV Booster capacity. The next Fermilab upgrade should include a replacement for the Booster [2]. The upgrade could be based on a continuation of the 800 MeV linac to ~2 GeV followed by either a Rapid Cycling Synchrotron (RCS) [3] or continuing the linac to 8 GeV. The 8 GeV linac would use PIP-II 650 MHz cryomodules followed by relatively inexpensive LCLS-II [4] or ILC-style cryomodules that use 1300 MHz SRF cavities, which have already been designed and mass-produced. This approach has been documented in a Snowmass 2021 white paper [5]. In this paper, we discuss the 8 GeV linac option. We begin with some discussion of the beam requirements and potential layouts for the linac. Constraints and features of the scenarios are then discussed.

## LINAC SCENARIO REQUIREMENTS

The Fermilab Proton Improvement Plan II (PIP-II) provides a new 800 MeV superconducting RF (SRF) linac that replaces the previous 400 MeV linac, enabling higher intensity injection into the Fermilab Booster and providing 800 MeV proton beam to other experiments. The primary purpose of PIP-II is to provide enhanced beam power delivery from the Main Injector to DUNE (Deep Underground Neutrino Experiment) [3]. This is enabled by increasing the beam energy and intensity delivered by the linac to the Fermilab Booster and increasing the Booster cycle rate. Table 1 shows high-level parameters of the Fermilab beam to DUNE before and after PIP-II, as presented in the Fermilab PIP-II Design Report. PIP-II increases the Booster cycle rate to 20 Hz and the beam intensity to $6.6 \times 10^{12}$ protons/pulse, enabling beam power of ~1 to 1.2 MW at beam energies of 60 to 120 GeV.

Further improvements will require replacement of the Booster with a higher-capacity injector. The initial design specification for the upgrade is that it should enable at least ~2.4 MW to DUNE from the MI [6]. High-level performance goals are presented in Table I.

Table 1: High-Level Parameters for PIP, PIP-II and the Booster Replacement Linac

| Parameter | PIP-I | PIP-II | BRL | Unit |
|---|---|---|---|---|
| Linac Energy | 400 | 800 | 8000 | MeV |
| Beam Current | 25 | 2 | 2 | mA |
| Pulse length | 0.03 | 0.54 | 2.1 | ms |
| Pulse Rep. Rate | 15 | 20 | 20 | Hz |
| Protons/pulse | 4.2 | 6.5 | 26.0 | $10^{12}$ |
| 8 GeV power | 80 | 166 | 700 | kW |
| Power to MI | 50 | 83-142 | 176-300 | kW |
| MI protons/pulse | 4.9 | 7.5 | 15.6 | $10^{13}$ |
| MI cycle time (120 GeV) | 1.5 | 1.2 | 1.2 | s |
| MI Power to DUNE | 0.7 | 1.2 | 2.5 | MW |
| 8 GeV other users | 30 | 83 | 500 | kW |

## LAYOUT

The BRL must take beam from the PIP-II linac into the MI/RR. The configuration is constrained by the fixed location of PIP-II and its proximity to the MI. Figure 1 shows a possible scenario. The PIP-II linac is extended to ~1 GeV by adding 2 cryomodules within the lattice at the end of the PIP-II tunnel, using drift spaces reserved for future extensions. The beam exiting that linac is bent at ~45° into a 1→2.4 GeV linac, which uses ~10—12 PIP-II 650 MHz cryomodules, requiring ~120—150 m. The total length available is ~290 m; the additional length will be used for optics matching and collimation. The transition energy depends on future design optimizations and applications; we consider 2.4 GeV as an initial choice.

The beam then goes through an achromatic bend of approximately 105° to be pointed toward injection into the Recycler Ring (RR) at MI-10. The following ~500 m transport includes a 2.4→8 GeV pulsed linac, consisting of LCLS-II-style 1300 MHz cryomodules [4], and takes the beam toward the MI at MI-10. The 2.4→8 GeV pulsed linac requires ~20 cryomodules, which occupy ~250 m. The facility would include transfer lines for intensity frontier experiments at ~1 GeV, 2.4 GeV and 8 GeV.

___________
*Work supported by Fermi Research Alliance, LLC under Contract No. DE-AC02-07CH11359 with the U.S. Department of Energy, Office of High Energy Physics.
† neuffer@fnal.gov

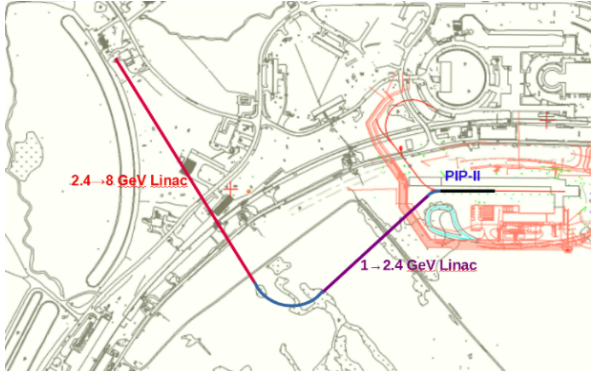

Figure 1: Layout of the 8 GeV linac from PIP-II to the MI, with MI-10 injection.

## SRF COMPONENTS

The BRL will contain 650 MHz and 1300 MHz cryomodules. Recent improvements in cavity performance by nitrogen doping and low-T cavity bake will be incorporated into the designs [7-9]. Parameters of the resulting SRF systems are presented in Table 2. Figure 2 shows cross-sections of the cryomodules, displaying cavities, couplers, and other components. At the parameters of Table 2, ~11 650-MHz cryomodules and ~20 1300-MHz cryomodules are required for the linac.

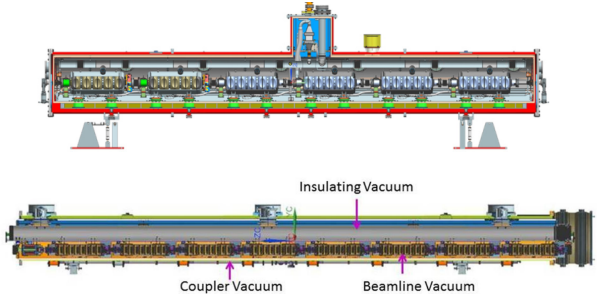

Figure 2: Cross-sections of 650 MHz (above) and 1300 MHz cryomodules (below), showing cavities and other components [1, 4].

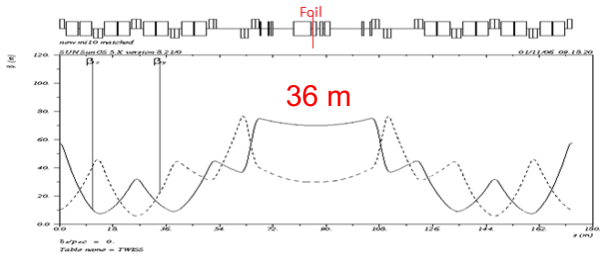

Figure 3: Layout with betatron functions for the RR-10 straight section, adapted for foil injection. A ~36 m segment between focusing quads is reserved for injection kickers, injection bump magnets and stripping foil.

Table 2: SRF Parameters

| Parameter | 650 MHz | 1300 MHz |
|---|---|---|
| $\beta$ (v/c) | 0.9 | 1.0 |
| Cells/cavity | 5 | 9 |
| Cavity length | 1.04 m | 1.038 m |
| R/Q | 638 $\Omega$ | 1036 $\Omega$ |
| $G=Q_0 R_s$ | 255 $\Omega$ | 270 $\Omega$ |
| Gradient $E_{acc}$ | 21 MV/m | 35 MV/m |
| $E_{max}$ | 46.8 MV/m | 70 MV/m |
| $B_{max}$ | 88 mT | 150 mT |
| $Q_0$ | $6.0 \times 10^{10}$ | $2.5 \times 10^{10}$ |
| $I_{H^-}$ current | 2-5 mA | 2-5 mA |
| $Q_L$ | $3.4 \times 10^7$ | $1.7 \times 10^7$ |
| Losses @2K | 16 W | 65 W |
| Cavity rf power | 120 kW | 184 kW |
| Cavities/cryo | 6 | 8 |
| Cryomodule L | 9.9 m | 12.5 m |
| Cavities needed | 66 | 160 |
| Cryomodules | ~11 | 20 |

## INJECTION

In Fig. 1, the 8 GeV H⁻ beam is directed toward MI-10, and injection into the MI or the RR in that region should be possible. However, the MI-10 straight section has been designated as the location for extraction of 120 GeV MI beam toward the LBNF target [10]. That extraction system precludes direct injection into the MI. Therefore, our baseline injection scenario is foil injection into RR-10. The straight section must be modified to include a large beam size at the foil (large $\beta_x$, $\beta_y$), and incorporate kickers and foils (see Fig. 3) [11].

High intensity multiturn H⁻ injection into the RR/MI, with injection painting and foil heating, was simulated by Drozhdin et al. [12] and further explored by Neuffer [13]. The preferred injection procedure is to split the injection into a few separate shorter injections, spaced by the pulsed linac rep. rate, following a foil painting program to minimize the number of foil hits. Figure 4 displays calculations of foil heating in a 6-step injection at 1, 2, and 4 mA currents, which reduces peak T to 2200, 1660, and 1250 °K, respectively. The 2 and 4 mA numbers are acceptable.

Foil damage is common in high-intensity H⁻ injection, and limits potential improvements. Laser assisted injection should be a future upgrade and will be relatively easy at 8 GeV [14]. The beam frame photon energy and power are magnified by factors of $\gamma(1+\beta\cos(\theta))$ and $\gamma^2(1+\beta\cos(\theta))^2$, respectively, and available high-power infrared lasers can be used.

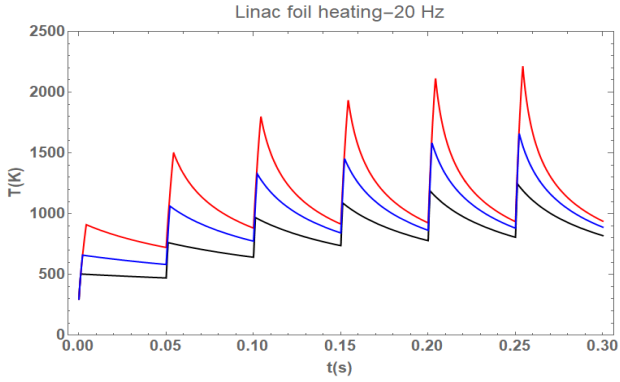

Figure 4: Foil heating in 6-step injection (4.2 mA-ms/step) at 1 (red), 2 (blue) and 4 mA (black).

## BEAM LOSSES

Beam losses can be a significant limitation. Important loss mechanisms include [15, 16]:

- Injected beam losses. A fraction of the beam misses the foil and some of the beam that hits the foil will not be fully stripped. Most of these should be captured by the injection absorber; uncontrolled losses should be less than ~0.2%;
- Magnetic stripping of H$^-$. Fields are kept small enough to keep losses less than ~$10^{-8}$/m, which requires B(T)×P(GeV/c) < 0.48, where B is the bending field and P is the H$^-$ momentum;
- Intrabeam scattering and stripping;
- Black body radiation stripping.

The injection region will be designed with a beam dump that captures most of the injection beam losses; this will confine activation to a limited area.

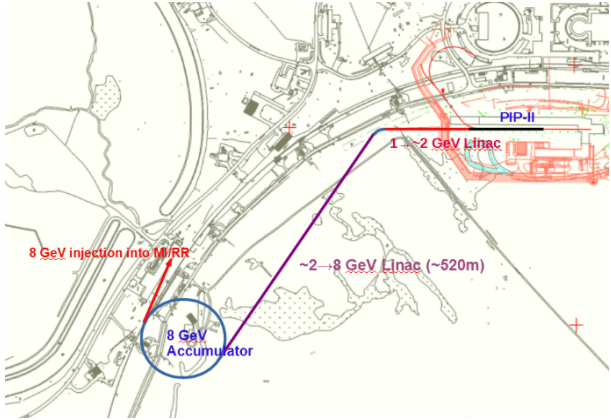

Figure 5: Layout of the 8 GeV linac with a new accumulator ring, with injection at MI 60 or 62.

## VARIANT SCENARIO

MI-10 extraction toward DUNE makes it difficult to inject H$^-$ into RR-10 and the injection will lead to losses in the MI enclosure that are difficult to mitigate – the MI is already loss-limited. An alternative layout is shown in Fig. 5. In this scenario the linac beam is injected into a new ~8 GeV Accumulator Ring, which is placed within the old Main Ring circle, so that the stored proton beam can then be injected into MI-60 or MI-62. [17, 18] The PIP-II linac is extended to ~2 GeV (using ~10 650-MHz cryomodules), where a bend of ~45° directs the beam into a new 2-8 GeV linac composed of 1300 MHz cryomodules, and the linac transport directs ~8 GeV beam into an accumulator ring. The accumulator would be an ~480 m circumference ring with a race-track shape, with ~100 m long straight sections, optimized for foil injection. Six pulses (each ~4.2 mA-ms) would be accumulated and transferred into the MI/RR by "box-car" stacking and then accelerated in the MI for DUNE. This geometry avoids interference with MI extraction and adds a storage ring that could feed other experiments with 8 GeV beam. Note that if injection is directly into the MI, the energy can be modified from 8 GeV for operational flexibility.

A lattice for the 8 GeV linac has been developed and simulated, with results presented in Fig. 7. In this scenario gradients were limited to 30 MV/m, and a 26 cryomodule lattice (325 m) is obtained and this fits easily within the ~520 m linac slot. The remaining space is used for matching optics, particularly for matching into Accumulator injection. An Accumulator lattice is under development.

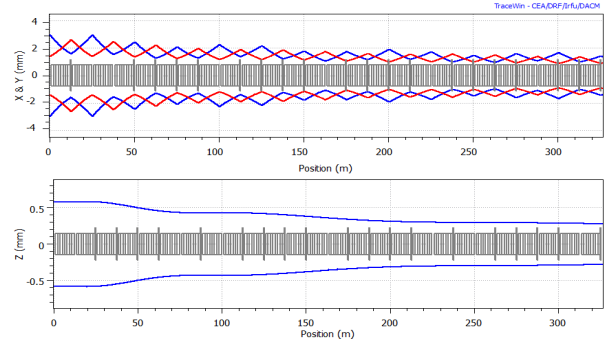

Figure 7: 2→8 GeV linac lattice with beam sizes, for $E_{acc}$ limited to less than 30 MV/m, 26 cryomodules (325 m).

## DISCUSSION AND FUTURE DIRECTIONS

The 8 GeV linac capitalizes on Fermilab experience and improvements in SRF, directly using cavity and cryomodules as developed for LCLS-II. The key production facilities already exist, including infrastructure for fabrication and testing. Construction will therefore be relatively cost-effective with low technical and cost risk.

There is substantial capability for power extension. For example, 20 Hz pulsed operation at 5 mA, 5 ms injection, would provide 2 MW power at 8 GeV, possibly enough for ν-factory or μ$^+$-μ$^-$ Collider operation. CW operation may enable even more power.

R&D will be needed to develop this into a complete project. The continuing SRF research should add further improvements. R&D focused on laser injection would dramatically enable higher performance. The many possible applications of this new high-intensity facility must be identified.

# REFERENCES

[1] M. Ball *et al.*, "The PIP-II Conceptual Design Report", Report No. FERMILAB-TM-2649-AD-APC, Fermilab, Batavia, IL, USA, 2017. `doi:10.2172/1346823`

[2] S. Holmes *et al.*, "Project X Accelerator Reference Design, Physics Opportunities, Broader Impacts", Report No. FERMILAB-TM-2557, Fermilab, Batavia, IL, USA, 2013. `doi:10.2172/1083535`

[3] R. Ainsworth *et al.*, "An Upgrade Path for the Fermilab Accelerator Complex", Report No. TM-2754-AD-APC-PIP2-TD, Fermilab, Batavia, IL, USA, 2021. `doi:10.2172/186535`

[4] J. Stohr, "Linac Coherent Light Source II (LCLS-II) Conceptual Design Report", SLAC Report No. SLAC-R-978, SLAC National Accelerator Laboratory, Menlo Park, CA, USA, 2019. `doi:10.2172/1029479`

[5] S. Belomestnykh *et al.*, "An 8 GeV Linac as the Booster Replacement in the Fermilab Power Upgrade", Report No. FERMILAB-FN-1131-AD-TD, Fermilab, Batavia, IL, USA, 2022. `doi:10.18429/JACoW-IPAC2021-MOPAB190`

[6] R. Acciarra *et al.*, "Long-Baseline Neutrino Facility (LBNF) and Deep Underground Neutrino Experiment (DUNE) Conceptual Design Report Volume 1: The LBNF and DUNE Projects", 2016. `doi:10.48550/arXiv.1601.05471`

[7] A. Grassellino *et al.*, "Unprecedented quality factors at accelerating gradients up to 45 MV/m in niobium superconducting resonators via low temperature nitrogen infusion", *Semicond. Sci. Technol.*, vol. 30, p. 094004, 2017. `doi:10.1088/1361-6668/aa7afe`

[8] A. Romanenko *et al.*, "Dependence of the residual surface resistance of superconducting radio frequency cavities on the cooling dynamics around $T_c$", *J. Appl. Phys.*, vol. 115, p. 184903, 2014. `doi:10.1063/1.4875655`

[9] D. Bafia *et al.*, "Gradients of 50 MV/m in TESLA Shaped Cavities via Modified Low Temperature Bake", in *Proc. SRF'19*, Dresden, Germany, Jun. 2019, pp. 586-591. `doi:10.18429/JACoW-SRF2019-TUP061`

[10] J. Strait *et al.*, "Long-Baseline Neutrino Facility (LBNF) and Deep Underground Neutrino Experiment (DUNE) Conceptual Design Report Volume 3: Long-Baseline Neutrino Facility for DUNE June 24, 2015", 2016. `doi:10.48550/arXiv.1601.05823`

[11] D. Johnson, "Conceptual Design Report of 8 GeV H- Transport and Injection for the Fermilab Proton Driver", *Beams-Doc 2597 v3*, Fermilab, Batavia IL, USA, 2007.

[12] A. I. Drozhdin *et al.*, "Modeling multiturn stripping injection and foil heating for high intensity proton drivers", *Phys. Rev. ST Accel. Beams*, vol. 15, p. 011002, 2012. `doi:10.1103/PhysRevSTAB.15.011002`

[13] D. Neuffer, "Injection Considerations for the PIP-III Linac to Main Injector", Report No. FERMILAB-FN-1099-AD, Fermilab, Batavia IL, USA, 2020. `doi:10.2172/1661676`

[14] S. Cousineau *et al.*, "First demonstration of laser-assisted charge exchange for microsecond duration H- beams", *Phys. Rev. Lett.,* vol. 118, p. 078401, 2017. `doi:10.1103/PhysRevLett.118.074801`

[15] D. Neuffer, "Beam Losses for the PIP-III Linac to Recycler/Main Injector", Report No. FERMILAB-TM-2472-AD, Fermilab, Batavia IL, USA, 2020. `doi:10.2172/1670446`

[16] J.-P. Carneiro *et al.*, "Numerical Simulations of Stripping Effects in High Intensity Hydrogen Ion Linacs", *Phys. Rev. ST Accel. Beams*, vol. 12, p. 040102, 2009. `doi:10.1103/PhysRevSTAB.12.040102`

[17] D. Neuffer, "Comment on the PIP-III RCS Location", *Beams-Doc 8874*, Fermilab, Batavia, IL, USA, 2020.

[18] S. Nagaitsev, private communication, 2022.